\documentstyle[12pt,aaspp4]{article}

\newcommand{\delt}{\partial_t}
\newcommand{\vl}{v_\ell}
\newcommand{\vlam}{v_\lambda}
\newcommand{\valf}{v_A}
\newcommand{\kz}{k_z}
\newcommand{\kperp}{k_\perp}
\newcommand{\ka}{{\bf k}_1}
\newcommand{\kb}{{\bf k}_2}
\newcommand{\kc}{{\bf k}_3}
\newcommand{\kd}{{\bf k}_4}

\newcommand{\cL}{{\cal L}}
\newcommand{\bx}{{\bf x}}
\newcommand{\ba}{{\bf a}}
\newcommand{\bap}{{\ba^\prime}}
\newcommand{\baperp}{\ba_\perp}
\newcommand{\al}{a_\parallel}
\newcommand{\bDelta}{{\Delta\boldeta}}

\newcommand{\sa}{ sA}
\newcommand{\pa}{ pA}
\newcommand{\boldeta}{\mbox{\boldmath $\eta$}}
\newcommand{\bxi}{\mbox{\boldmath $\xi$}}
\newcommand{\bnabla}{\mbox{\boldmath $\nabla$}}
\newcommand{\bzeta}{\mbox{\boldmath $\zeta$}}

\begin{document}

\lefthead{MHD Turbulence}

\righthead{P.~Goldreich and S.~Sridhar}

\title{MHD Turbulence Revisited}

\author{P.~Goldreich~$^1$ and S.~Sridhar~$^2$}

\affil{$^1$~California Institute of Technology\\
Pasadena, CA~91125, USA}

\affil{$^2$~Inter-University Centre for Astronomy and Astrophysics\\
Ganeshkhind, Pune 411 007, INDIA}

\authoremail{pmg@nicholas.caltech.edu and sridhar@iucaa.ernet.in}

\begin{abstract}
Kraichnan~(1965) proposed that  MHD turbulence occurs as a result of 
collisions between oppositely directed Alfv\'en wave packets.
Recent work has generated some controversy over the nature of non linear couplings between colliding Alfv\'en waves. 
We find that the resolution to much of the
confusion lies in the existence of a new type of turbulence, {\sl intermediate
turbulence}, in which the cascade of energy in the inertial range exhibits
properties intermediate between those of weak and strong turbulent 
cascades. Some properties of intermediate MHD turbulence are:
(i) in common with weak turbulent cascades,
wave packets belonging to the 
inertial range are long lived; (ii) however, components of the strain
tensor are so large that, similar to the 
situation in strong turbulence, perturbation theory
is not applicable; (iii) the breakdown of perturbation
theory results from the divergence of neighboring field lines due to
wave packets whose perturbations in velocity and magnetic fields are
localized, but whose perturbations in displacement are not; (iv)
3--wave interactions dominate individual collisions between wave packets,
but interactions of all orders $n\geq 3$ make comparable contributions to the intermediate turbulent energy cascade; (v) successive collisions are
correlated since wave packets are distorted as they follow diverging
field lines; (vi) in common with the weak MHD cascade, 
there is no parallel cascade of energy, and the cascade to small perpendicular
scales strengthens as it reaches higher wave numbers; (vii) For an appropriate weak excitation, there is a natural progression from a weak, through an intermediate, to a strong cascade.       
\end{abstract}


\section{INTRODUCTION}

There appears to be some consensus regarding the notion that turbulence in the 
ionized interstellar medium is of magnetohydrodynamic (MHD) origin.
However, even three decades after Iroshnikov~(1963)
and Kraichnan~(1965) first, independently, presented their ideas on MHD turbulence
(in an incompressible, highly conducting fluid), there is much debate 
on what this theory
really is. Recent controversy centers on the 
existence, or otherwise, of certain nonlinear interactions among 
Alfv\'en waves.
Sridhar \& Goldreich~(1994; hereafter SG) have 
argued that there are no resonant 3--wave 
interactions in weak MHD turbulence. They also asserted that the 
Iroshnikov--Kraichnan (IK) theory is incorrect, and constructed a 
theory of weak MHD turbulence based
on resonant 4--wave interactions; in this theory, non linear
interactions strengthen on small spatial
scales, resulting ultimately in a strong cascade proposed by  
Goldreich \& Sridhar~(1995; hereafter GS). However, 
Montgomery \& Matthaeus~(1995) have maintained  that resonant 3--wave interactions 
are non empty, holding these  responsible for the anisotropic
cascade seen in the numerical simulations of Shebalin, Matthaeus, \& Montgomery~(1983). Moreover, Ng \& Bhattacharjee~(1996; hereafter NB) 
have recently presented
calculations which show that small amplitude wave packets do interact via
3--waves, so long as the $k_z=0$ ($\hat{z}$ is the direction of the local,
mean magnetic field) Fourier components of velocity and magnetic field perturbations are non zero. The present investigation is an 
analysis and resolution of this controversy. 

We consider magnetic turbulence in an incompressible fluid, 
although we simply call it MHD turbulence.
When the density and transport properties of the fluid 
are constant, the equations of 
(incompressible) MHD read

\begin{eqnarray}
\delt{\bf b}&=&\bnabla{\bf\times}({\bf v\times b})+
\kappa\bnabla^2{\bf b}, \nonumber \\
\delt{\bf v}&=&-({\bf v}\cdot\bnabla){\bf v}+
({\bf b}\cdot\bnabla){\bf b}-\bnabla p +\gamma\bnabla^2{\bf v}, \\
&{}&\bnabla\cdot{\bf v}=\bnabla\cdot{\bf b}=0, \nonumber 
\label{incompress}
\end{eqnarray}
\noindent where ${\bf v}$ is the velocity,
${\bf b}={\bf B}/\sqrt{4\pi\rho}$ is the magnetic field in
velocity units, and $p$ is the ratio of total pressure to the density.
The dissipation provided by $\kappa$
and $\gamma$ is important only on small spatial scales, so 
we ignore the dissipative terms. 
The homogeneous state, ${\bf v}_0=0$, ${\bf B}_0=B_0\hat{z}$, is a
stable, static solution of equations~(\ref{incompress}). 
{\sl Shear} Alfv\'en  waves and
{\sl pseudo} Alfv\'en waves are the two  kinds of linear
perturbations about this equilibrium, the latter being the incompressible
limit of the slow magnetosonic wave. Both kinds of waves have the
same dispersion relation, namely,
$\omega=\valf|\kz|$, where $\valf=B_0/\sqrt{4\pi\rho}$ is called
the Alfv\'en speed.
The perturbed velocity and magnetic fields are related by
$\delta{\bf v}=\pm\delta{\bf b}$, where the upper/lower signs correspond to
waves traveling antiparallel/parallel to ${\bf B}_0$ (with $\kz<0$
and $\kz>0$ respectively).
Equations~(\ref{incompress}), with $\kappa=\gamma=0$, 
possess the remarkable property of 
allowing for nonlinear generalizations of the linear Alfv\'en waves.
Mutual cancellation of nonlinear terms permits the following 
wide class of exact solutions:
if $\delta{\bf v}({\bf x})=-\delta{\bf b}({\bf x})$ at some instant
of time, $t=0$, it can be checked that
$\delta{\bf v}(x, y, z -\valf t)=-\delta{\bf b}(x, y, z-\valf t)$
for all time, irrespective of the functional form of
$\delta{\bf v}({\bf x})$ (c.f. Parker~1979). This nonlinear solution
describes a wave packet
of arbitrary form traveling nondispersively in the direction of
${\bf B}_0$. Similarly, we can also construct another class of nonlinear
solutions, with $\delta{\bf v}=\delta{\bf b}$, that travels nondispersively
in a direction opposite to ${\bf B}_0$. Both types of nonlinear solutions
are stable, and the dynamics is simple so long as there is no spatial
overlap (``collisions'') between oppositely moving wave packets.

Let us consider a  situation that is less restrictive than perturbations
about a static equilibrium  (a uniform magnetic field). We imagine that
turbulent motions on a large scale are set up in the fluid by stirring it 
in a random but statistically steady fashion. These create large--scale,
disordered velocity and magnetic fields. Kinetic and magnetic
energies will cascade to smaller spatial scales.\footnote{On scales
small enough for the $\kappa\bnabla^2{\bf b}$
and the $\gamma\bnabla^2{\bf v}$ terms to be important, kinetic and magnetic
energies will dissipate into heat.}~ The central problem of MHD turbulence
is to determine the statistical steady--state amplitudes of the fluctuations in ${\bf v}$ and ${\bf b}$, on intermediate spatial scales---the so called 
inertial--range power spectrum. Mathematically speaking, this cascade 
through spatial scales should emerge from the effect of
the nonlinear terms in equations~(\ref{incompress}). Physically, it is
common to speak of interactions between ``eddies'' 
(c.f. Frisch~1995). Do interactions between eddies of dissimilar
sizes make significant contributions to the form of the cascade?
For hydrodynamic turbulence, the answer seems to be, ``no, the 
dominant interactions are between eddies of similar spatial scales''.
The reason for such a locality in interactions between eddies is that
the sweeping due to the velocity field of a large sized eddy (on a smaller
eddy) may be transformed away by a local Galilean transformation. 
{\sl Kraichnan noted that, for MHD turbulence, such a transformation has 
no effect at all on the magnetic field of the large eddy; the magnetic
field of a large eddy acts upon smaller eddies in much the same manner as
the mean magnetic field does on Alfv\'en wave packets. Hence MHD turbulence
in an incompressible fluid should reduce to turbulence of interacting 
Alfv\'en wave packets}. We recall from the previous paragraph that finite 
amplitude wave packets that travel in the positive/negative $z$--directions
do so without change in form, so long as oppositely directed packets
do not overlap. {\sl Kraichnan realized that the cascade of energy in
MHD turbulence occurs as a result of collisions between 
oppositely directed Alfv\'en wave packets}. 

To derive the results of the IK theory, consider a statistically steady, 
isotropic excitation of amplitude $\vl\ll \valf$, on outer scale $\ell$, 
of the static equilibrium mentioned earlier. Alfv\'en waves have $\delta v\sim\delta b$, so that $\vl$ is the amplitude of excitation of both velocity and magnetic fields. The resulting fluctuations may,
at any time, be decomposed into Alfv\'en wave packets with scales $\lambda\lesssim \ell$ traveling in the positive
and negative $z$ directions.  IK assumed that the energy transfer is 
{\sl local} and {\sl isotropic} in ${\bf k}$--space. Collisions between oppositely directed wave packets
occur over times of order $\omega_k^{-1}\sim (\valf k)^{-1}$, and these
create small distortions. During one collision, each wave packet suffers
a fractional perturbation

\begin{equation}
{\delta \vlam\over \vlam}\sim {d\vlam\over dt}(\valf k\vlam)^{-1}
\sim {\vlam\over\valf}\ll 1\,.\label{ikpert}\end{equation}
\noindent During successive collisions, these perturbations add with
random phases. The number of collisions for the fractional perturbations
to build up to order unity is

\begin{equation}
N_\lambda\sim \left({\vlam\over\delta\vlam}\right)^2\sim
\left({\valf\over\vlam}\right)^2\,.\label{iknumber}
\end{equation}
\noindent The energy cascade rate is $\epsilon\sim \vlam^2/t_\lambda$, where 
the cascade time is given by $t_\lambda\sim N_\lambda/(\valf k)
\sim \valf/(k\vlam^2)$. The three--dimensional energy spectrum, $E(k)$,
is related to the velocity fluctuation by $\vlam^2\sim k^3E(k)$. 
Using Kolmogorov's hypothesis of the scale independence of $\epsilon$, 
we obtain the following scalings for the inertial--range of the IK
theory:

\begin{equation}
{\vlam\over\vl}\sim \left({\lambda\over\ell}\right)^{1/4}\,\qquad\qquad 
E(k)\sim {\vl^2\over\ell^{1/2}k^{7/2}}\,.\label{ikspectrum}
\end{equation}
\noindent With these scalings, 

\begin{equation}
N_\lambda\sim \left({\valf\over\vl}\right)^2\left({\ell\over\lambda}\right)^{1/2}\, 
, \label{iknumberp}
\end{equation}

\noindent so the cascade weakens as $\lambda$ decreases.

\section{INTERMEDIATE TURBULENCE}
\subsection{A New Cascade Based on 3--Wave
Interactions}

SG argued that the IK theory, although
seemingly plausible, was basically incorrect. 
Here we discuss the reason for the failure of this theory, and propose a
new cascade based on 3--wave interactions. Let us begin by listing 
three key features of the IK theory.

\noindent (i)  Wave packets of size $\lambda$ live for $N_\lambda$ wave periods;
large values of $N_\lambda$ correspond to weak interactions between wave
packets of size $\lambda$. From equations~(\ref{iknumber}) and
(\ref{ikspectrum}),  
it may be verified that 
$N_\lambda\propto 1/\lambda^{1/2}$, so that the cascade weakens as it
progresses into the inertial--range.

\noindent (ii) From equation~(\ref{ikpert}), the 
fractional perturbation suffered by a wave packet during one 
collision is $\sim (\vlam/\valf)\ll 1$; In the IK theory, interactions
between wave packets are described by the lowest order non linear terms.

\noindent (iii) Isotropy is assumed in the derivation of the IK theory.
 
\noindent The derivation of this theory is essentially 
heuristic, roughly as given above in equations~(\ref{ikpert})---(\ref{ikspectrum}).
We note this, so that it is clear to the reader that there isn't a more 
rigorous version of the ``theory'' that we happened not to mention.
Together, features (i) and (ii) imply that the ``elementary interactions''
between Alfv\'en waves must satisfy the 3--wave resonance conditions:
 
\begin{equation}
\ka +\kb=\kc\,,\qquad\qquad
 \omega_1 +\omega_2=\omega_3
\,,\label{resthree}
\end{equation}

\noindent where $\omega_k=\valf|k_z|\,$. 
Shebalin, Montgomery \& Matthaeus~(1983) noted that
the only non trivial solutions of  equations~(\ref{resthree}) require that
one of either $k_{1z}$, or $k_{2z}$ must be zero. This implies that
waves with values of $k_z$ not present initially cannot be created
during collisions between oppositely directed wave packets. As they 
point out, there is no parallel (i.e. along $k_z$) cascade of energy.
Thus the turbulence must be anisotropic, and energy should cascade to large
$\kperp$.

In what follows, we
derive an anisotropic version of IK theory. As before, 
we imagine that the system is stirred in a statistically steady and
isotropic fashion
such that $\vl\ll\valf$ on outer scale $\ell$.
The absence of a parallel cascade implies that wave packets belonging to 
the inertial range have parallel scales $\ell$ and perpendicular scales $\lambda\ll\ell$. We estimate the spectrum of the anisotropic cascade by
modifying the arguments of equations~(\ref{ikpert})---(\ref{ikspectrum})
so as to
keep track of parallel, and perpendicular scales.
In one collision between two such oppositely directed
wave packets, the fractional perturbation is given by

\begin{equation}
{\delta\vlam\over\vlam}\sim {d\vlam\over dt}(\valf\kz\vlam)^{-1}\sim
{\ell\vlam\over\lambda\valf}\ll 1\,.\label{gspert}
\end{equation}

Adding up the perturbations due to successive collisions with random phases,
the number of collisions over which the perturbations grow to order unity is

\begin{equation}
N_\lambda\sim \left({\vlam\over\delta\vlam}\right)^2\sim
\left({\lambda\valf\over\ell\vlam}\right)^2\,.\label{gsnumber}
\end{equation}

\noindent Making use of the steady-state relation (``Kolmogorov's hypothesis'')

\begin{equation}
{v_\lambda^2\over t_\lambda}\sim {v_\ell^2\over t_\ell}\sim\epsilon\,,
\end{equation}

\noindent we obtain scaling relations for the velocity fluctuations, as well 
as the three dimensional energy spectrum of the anisotropic, 3--wave 
cascade

\begin{equation}
{v_\lambda\over\vl}\sim\left({\lambda\over\ell}\right)^{1/2}\,,\qquad\qquad
E(k_z, \kperp)\sim {\vl^2\over \kperp^3}\, .\label{gsspectrum}
\end{equation}

\noindent It follows that 

\begin{equation}
N_\lambda\sim \left({\valf\over\vl}\right)^2{\lambda\over\ell}\, ;
\label{gsnumberp}
\end{equation}

\noindent so nonlinearity increases along the inertial range
of the new cascade. 

The anisotropic cascade differs from the original IK cascade by: (i) being
anisotropic; (ii) having a different spectrum; (iii) strengthening at high wave
number. This final difference has a profound consequence. It implies that
the anisotropic cascade has a limited inertial range, thereby
diminishing its applicability in astronomical contexts where the 
excitation at the outer scale is likely to be quite strong. 

It turns out that this 
anisotropic cascade is an example of a new type of turbulence, 
which we call {\sl Intermediate Turbulence}, because it has properties
intermediate between those of weak and strong turbulence. In particular, intermediate turbulence does not submit to perturbation theory. 

\subsection{ The Failure of Perturbation Theory
in Intermediate Turbulence}

\subsubsection{Field Line Geometry}

To lowest order in perturbation theory, wave packets move along field lines.
Thus the breakdown of perturbation theory may be understood physically
by studying the geometry of the divergence of a bundle of field lines.
Assume that the mean 
field lies along the $z$ axis. Consider wave packets localized in velocity and magnetic 
field perturbations, but not in displacement, having longitudinal
scale $\ell$ and transverse scale $\lambda$ with 
$\lambda/\ell < 1$. Let us require that the velocity fluctuation, $\vlam$,
is small enough, 

\begin{equation}
\chi\equiv {\ell v_\lambda\over \lambda \valf}\ll 1, 
\label{eqchi}\end{equation}

\noindent so that $N_\lambda\ll 1$ (cf. eqn. [\ref{gsnumber}]). The rms 
differential inclination of the local
field across scale $\lambda$ is $\theta_\lambda\sim v_\lambda/v_A$. It is correlated over distances $\ell$ and $\lambda$ parallel and orthogonal to $z$. 

Let us focus on a pair of neighboring field lines separated by 
$\lambda$ at $z=0$. The expectation value of the separation 
between these field lines, $\Delta$, varies such that 

\begin{equation}
\Delta^2\sim \lambda^2 + \theta_\lambda^2\ell|z|,\label{eqtheta}
\end{equation}
\noindent for $|z|\gg \ell$. The distance along
$z$, over which $|\Delta|$ increases by a factor of order unity 
from its initial value of $\lambda$ at $z=0$, is 

\begin{equation}
L_*\sim \left({\lambda v_A\over \ell v_\lambda}\right)^2\ell
\sim {\ell\over\chi^2}\, ;\label{lstar}
\end{equation}

\noindent $L_*$ is an function of $\lambda$. The significance of $L_*$ in 
intermediate turbulence follows because turbulence involves
the transfer of energy across scales. If $\chi\ll 1$, single 
interactions between wave packets result in small perturbations.
Cascading of energy requires of order 

\begin{equation}
N_\lambda\sim{1\over\chi^2}\sim {L_*\over \ell}\gg 1\label{eqN}
\end{equation}

\noindent such interactions. 

\subsubsection{Nonlinear Interactions in Intermediate Turbulence}

We have deduced the form of the spectrum of intermediate MHD turbulence 
from scaling arguments based on 3-wave couplings. Moreover, these
interactions are weak in the sense that $N_\lambda\gg 1$. This might suggest
that 3-wave interactions dominate those of higher order and that
a rigorous derivation of the steady-state cascade might result from 
truncation at this order. Unfortunately, this is incorrect; interactions 
of all orders have similar strengths. 

Consider the distortion suffered during a single collision between
oppositely directly wave packets of similar strength having perpendicular and parallel dimensions $\lambda\lesssim \ell$ and
$\ell$. We assume that $\ell v_\lambda\ll \lambda \valf$. Contributions
from interactions involving n-waves may be written as

\begin{equation}
{\delta^n v_\lambda\over v_\lambda} \sim \left({\ell v_\lambda\over
\lambda \valf}\right)^{n-2}. \label{eqdeln}
\end{equation}

\noindent Clearly 3-wave interactions dominate those of higher order for
{\sl individual collisions}. 

Next we consider the cumulative distortion due to n-wave interactions as
a wave packet travels a distance $\ell\ll z\ll L_*$. We can picture the distortion as arising from the shearing of the packet as it follows
the differential wandering of neighboring field lines.\footnote{A uniform displacement of a wave packet does not contribute to the energy cascade.}~
The net displacement of individual field lines over distance $z$ defines
a vector field whose shear tensor transforms the packet's shape.\footnote{This transformation is subject to the constraints of fluid incompressibility and magnetic flux freezing.}~ For $\ell\ll z\ll L_*$ this transformation
is close to the identity, so it may be expanded in a Taylor series.
The expansion parameter is $(z/L_*)^{1/2}$, the dimensionless measure of
fractional spreading over distance $z$ of a bundle of field lines whose cross sectional radius at $z=0$ is $\lambda$. 
Terms of order $n-2$ in $v_\lambda$ correspond to $n$-wave interactions.\footnote{Since wave packets follow field lines only
to lowest order, nonkinematic terms appear at orders $n\geq 4$. Those of
$n=4$ are given in \S 3.4.}~ These
terms have the form

\begin{equation}
{\delta^n v_\lambda\over v_\lambda} \sim \left({\ell v_\lambda\over
\lambda \valf}\right)^{n-2}\left({z\over\ell}\right)^{(n-2)/2}\sim 
\left({z\over L_*}\right)^{(n-2)/2}. \label{eqdelnL}
\end{equation}

\noindent A notable feature of equation~(\ref{eqdelnL}) is that contributions
from higher order ($n\geq 4$) interactions carry extra factors of $(z/\ell)^{1/2}$. By
extra factors we mean those beyond the single factor $(z/\ell)^{1/2}$
expected to arise from the addition of a sequence of
independent interactions between pairs of wave packets. These factors 
reveal an interdependence among collisions associated with the nonlocalized
field line displacements of the wave packets.\footnote{These extra
factors of $(z/\ell)^{1/2}$ are absent in the corresponding formula for wave
packets which are localized in displacement.}  

Intermediate turbulence is nonperturbative because distortions of all orders become large as $z\to L_*$. This is as expected because the cascade time across scale $\lambda$ is $t_\lambda\sim L_*/\valf$. It implies that n-wave
interactions of all orders $n\geq 3$ make comparable contributions to the
energy cascade. 

\subsubsection{Lagrangian Perturbation Theory}

A fluid element whose 
Lagrangian coordinate is $\ba$ has Eulerian location $\bx$, at time $t$,
given by 

\begin{equation}
\bx = \ba + \bxi(\ba, t), \label{disp}
\end{equation}

\noindent where $\bxi(\ba,t)$  is the displacement field. 
Velocity and magnetic field perturbations at the Eulerian location are
defined in terms of the displacement vector by

\begin{equation}
{\bf v}(\bx, t)={\partial\bxi\over\partial t}(\ba, t)\,,\qquad\qquad
{\bf b}(\bx, t)=\valf{\partial\bxi\over\partial \al}(\ba, t)\,.
\label{vbdef}
\end{equation}

\noindent SG employed a formulation of the MHD action
due to Newcomb~(1962), and developed a 
Lagrangian perturbation theory for  weak MHD turbulence. The Lagrangian
is a functional of the strain tensor field, in other words, the gradient 
of the displacement vector field. Expansion of the Lagragian density in
powers of the strain tensor yields terms of second order, $n=2$, and 
then fourth and higher orders, $n\geq 4$. The
absence of third order terms signifies that wave packets follow field lines
to lowest nonlinear order. The absence of Lagrangian perturbations based on
3-wave interactions implies that Eulerian perturbations due to 3-wave interactions are purely kinematic.  Kinematic contributions
to Eulerian perturbations are also present at each higher order, $n\geq 4$.  
But these are augmented by dynamic perturbations arising from order $n\geq 4$ 
terms in the expansion of the Lagrangian.    
  
As before, we consider small amplitude, $(\ell v_\lambda/\lambda\vl)\ll
1$ wave packets localized in ${\bf v}$ and ${\bf b}$ but not in
$\bxi$.  Roughly speaking, convergence of the perturbation expansion
requires the components of the strain tensor to be smaller than 
unity.\footnote{From this point on it is best to proceed in Fourier space, since
the breakdown of perturbation theory is closely related to the
behavior of the energy spectrum (i.e. velocity, or magnetic field
power spectra) at small $k_z$.}~  An ensemble of independent wave
packets generates an energy spectrum which is flat for $k_z\ell\lesssim
1$.  Since $\omega=v_A|k_z|$, the power spectrum of the displacement
vector field varies as $k_z^{-2}$ for $k_z\ell\ll 1$.  The same
behavior characterizes the power spectra of some of the components of
the strain tensor.  It implies that these components diverge. The
divergence is the mathematical expression of the spreading of field
line bundles described in \S 2.2. In this light, the failure of
perturbation theory is seen to be generic, and not just a consequence of an unfortunate choice of perturbation variable.

Next we investigate the effect of cutting off the energy spectrum below
$k_zL\sim 1$, where $L\gtrsim\ell$. The most strongly divergent components of
$\xi_{i,j}$ have power spectra given by 

\begin{equation}
|\tilde{\xi}_{i,j}({\bf k})|^2\sim \left({v_\lambda\over v_A}\right)^2
{\ell\over k_z^2}. \label{psgrad}
\end{equation} 

\noindent Multiplying equation (\ref{psgrad}) by $k_\perp^2$ and integrating from 
$k_z\sim L^{-1}$ to $k_z\sim \ell^{-1}$, we obtain the
average value of $|\xi_{i,j}|^2$ due to power in this wave number interval, 

\begin{equation}
\overline{|\xi_{i,j}|^2}\sim\lambda^{-2}\int^{\ell^{-1}}_{L^{-1}}dk_z\, |\tilde{\xi}_{i,j}({\bf k})|^2\sim  \left({\ell v_\lambda\over \lambda v_A}\right)^2 {L\over \ell}.\label{stensor}
\end{equation}

\noindent Thus

\begin{equation}
\overline{|\xi_{i,j}|^2}\sim {L\over L_*}\, ,
\label{strainmag}
\end{equation} 

\noindent where $L_*$ is defined in equation (\ref{lstar}). 
Once again we see the crucial role played by $L_*$; the perturbation expansion converges if the energy spectrum is cutoff below $k_zL_*\sim 1$.

The absence of third order terms in the expansion of the Lagrangian 
signifies the absence of resonant 3-wave interactions in perturbative
MHD turbulence. Weak MHD turbulence based on 4-wave
interactions is discussed in \S~3.2

\section{ON WEAK AND INTERMEDIATE TURBULENCE}
\subsection{Types of Turbulence}

We have discussed, at some length, the intermediate cascade. It is  
time to state in a precise manner the properties that characterize the three 
kinds of turbulence. Let us begin with a definition.
Weak turbulence is characterized by the following
properties:\footnote{For a standard discussion of these points see the
introduction to ``Kolmogorov Spectra of Turbulence I'' by Zakharov,
L'vov, and Falkovich, 1992.}

\noindent {\bf I}. Nonlinear interactions among an ensemble of waves which are weak in
the sense that the fractional change in wave amplitude during each
wave period is small.
 
\noindent {\bf II}. The existence of a convergent perturbation expansion for the nonlinear
interactions. Typically the small parameter is the fractional change in
wave amplitude during a wave period. 

\noindent When these two conditions are satisfied, a formal theory of {\sl resonant}
wave interactions may be derived. For the theory to be nonempty, either the
3--wave or 4--wave resonance relations must possess nontrivial solutions.
Power spectra of cascades arise as stationary
solutions of the kinetic equation describing modal energy transfer.

Specifically, for MHD turbulence, we find 
 
\noindent 1. The turbulent cascade based on 4--wave interactions, derived in SG, is
the unique {\sl weak} cascade that satisfies both I and II above.
Moreover, this cascade is realizable; it could in principle be set up
experimentally. While 3-wave interactions do not vanish, 
they are nonresonant and do not transfer energy among different waves.

\noindent 3. The critically balanced cascade described in GS is an example of
{\sl strong} turbulence. 
It violates both conditions I and II. The interaction time is of order the wave period. Interactions of all orders have comparable strengths; there is no valid perturbation expansion. 

\noindent 3. There is a third type of MHD turbulence that satisfies I, but not II.
It is an example of what we have called {\sl Intermediate Turbulence}. 
This turbulence exhibits weak interactions, but strains in the fluid are so strong that perturbation theory diverges. The 3--wave interactions
are included, but not dominant; it turns out that interactions of all
orders contribute equally weakly. The inertial range spectrum of 
intermediate MHD turbulence is given, for the first time, in 
equations~(\ref{gsspectrum}) of this paper.

\subsection{Weak Alfv\'enic Turbulence}

Having discussed intermediate turbulence in some detail, we provide 
a brief outline of the theory of weak MHD turbulence that SG constructed
using resonant 4--wave interactions. By studying the resonant terms of the fourth order Lagrangian, SG derived a formal kinetic equation for the
evolution of energies (more precisely, ``wave action'') in different modes,
and proved that a cascade of energy, indeed, emerges as a stationary solution. 
The elementary interactions involve scattering of two waves which respect the 
following conservation laws:

\begin{equation}
\ka +\kb=\kc + \kd\,,\qquad\qquad
 \omega_1 +\omega_2=\omega_3
+\omega_4\,.\label{reso4}
\end{equation}

\noindent Using $\omega_k=\valf|k_z|$, and the $z$ component of the equation involving
the ${\bf k}$'s, SG prove that $k_{1z}=k_{3z} >0\,$, and $k_{2z}=k_{4z} <0\,$.
Of course, the symmetry of equations~(\ref{reso4}) allows us to flip the signs of
all the $k_z$, or permute indices 3 and 4; the important point is that the 
scattering process described by equations~(\ref{reso4}) leaves the $k_z$ components
unaltered. This implies that waves with values of $k_z$ not present in the
external stirring, cannot be created by resonant 4--wave interactions. 

In addition to developing a formal theory, SG also provided a heuristic derivation of the weak 4--wave cascade for {\sl shear} Alfv\'en waves. 
Here we note the main properties of the weak cascade of shear 
Alfv\'en waves:

\noindent (i) As discussed above, there is no transfer of energy to small
spatial scales in the $z$ direction; the energy cascade in ${\bf k}$--space
occurs only along ${\bf k}_\perp$ (i.e in directions perpendicular to
the mean magnetic field).

\noindent (ii) The three dimensional energy spectrum, $E$, is defined by

\begin{equation}
\sum \vlam^2 =\int E(\kz, \kperp)\,{d^3k\over 8\pi^3}\,,
\label{defspec}
\end{equation}

\noindent where $\sum$ is a sum over wave packets of various scales.
Weak turbulence relies on a convergent perturbation theory. As discussed
in \S~2.2.3  this requires that the spectrum, $E$, be cut off for
$|\kz L_*|<1\,$. Moreover, weak 4--wave interactions do not change $\kz$, 
which implies that $E$
may have a quite arbitrary dependence on $|\kz|$. This simply depends 
on the nature of the excitation, and is not of much interest here. 
If $\lambda\sim\kperp^{-1}$ is a perpendicular
length scale belonging to the inertial--range, the scalings derived by SG for the weak, 4--wave cascade are

\begin{equation}
{\vlam\over v_\ell}\sim\left({\lambda\over\ell}\right)^{2/3}\,,\qquad\qquad
E(\kz, \kperp)\sim{v_\ell^2\over \ell^{1/3}\kperp^{10/3}}\, ,
\label{direct}
\end{equation}

\noindent (iii) The number of collisions needed for the
packet to lose memory of its initial state, 

\begin{equation}
N_\lambda\sim\left({\lambda v_A\over\ell\vlam}\right)^4\sim\left({v_A\over v_\ell}\right)^4\left({\lambda\over\ell}\right)^{4/3}\,\label{Nfour}
\end{equation}
 
\noindent Note that, in common with the spectrum of
intermediate turbulence (given in equation~\ref{gsspectrum}),
$N_\lambda$ decreases as the cascade proceeds to smaller $\lambda$.

\noindent (iv) In their treatment of weak turbulence, SG unwittingly made the 
assumption that $E$ was cut off at small $|\kz|$. While the weak turbulence 
of SG is realizable (see \S~4 later in this paper), this feature makes it
less applicable than intermediate turbulence.

\subsection{A Controversy, and its Resolution}

NB have proved, analytically, that 3--wave interactions 
between small amplitude wave packets are non zero if the $k_z=0$
Fourier components are non zero. On the other hand, SG have argued, using a
Lagrangian perturbation theory, that 3--wave interactions are 
absent in weak MHD turbulence. In \S~2 we claim that the interactions found by
NB lead to intermediate, rather than weak turbulence. Here we bolster that
claim by an explicit evaluation of 3-wave and 4-wave interactions in
Lagrangian coordinates. 

The simplest derivation employs the Lagrangian displacement vector
field as the basic variable (cf. eqns. [\ref{disp}] and [\ref{vbdef}]). 
Incompressibility implies that the transformation between Lagrangian and Eulerian coordinates have unit Jacobian. Thus

\begin{equation}
J=1+\bnabla\cdot\bxi-{1\over 2}\bnabla\bxi:\bnabla\bxi
+{1\over 2}\left(\bnabla\cdot\bxi\right)^2
+{1\over 3}\xi^{i,j}\xi^{j,k}\xi^{k,i}-{1\over 2}\left(\bnabla\cdot\bxi\right)
\xi^{i,j}\xi^{j,i}+{1\over 6}\left(\bnabla\cdot\xi\right)^3, \label{J}
\end{equation}

\noindent where $\bnabla$ refers to derivatives with 
respect to Lagrangian coordinates, and

\begin{equation}
\bnabla{\bxi}:\bnabla{\bxi}\equiv \xi^{i,j}\xi^{j,i}.
\end{equation}

The displacement vector $\bxi$ is split into transverse and longitudinal 
components, $\boldeta$ and $\bzeta$ respectively, such that 

\begin{equation}
\bxi=\boldeta+\bzeta, \label{split}
\end{equation}

\noindent with 

\begin{equation}
\bnabla\cdot\boldeta=0\,.\end{equation}

\noindent The components of $\boldeta$ are the two independent variables; 
$\bzeta$ is obtained from equation (\ref{J}). Thus

\begin{equation}
\bnabla\cdot\bzeta_2={1\over 2}\bnabla\boldeta_1:\bnabla\boldeta_1\, \label{delzeta2} \end{equation}

\noindent and 

\begin{equation}
\bnabla\cdot\bzeta_3= \bnabla\boldeta_1:\bnabla\boldeta_2 +\bnabla\boldeta_1:\bnabla\bzeta_2 -{1\over 3}\eta_1^{i,j}\eta_1^{j,k}\eta_1^{k,i} \, . \label{delzeta3}
\end{equation}

\noindent The first term on the right hand side of equation (\ref{delzeta3}) is included for completeness since $\boldeta_2=0$.\footnote{See equation (\ref{pertvel}) and the footnote which follows it. }

Following  the development in \S~3 of SG, we write the Lagrangian as

\begin{equation}
\cL={\rho\over 2}\int\, d^3a\left(\left|{\partial\bxi\over\partial 
t}\right|^2-\valf^2\left|{\partial\bxi\over \partial\al}\right|^2\right).
\label{lagr}
\end{equation}

\noindent However, we present our calculations in real space, rather than 
in Fourier space; the results are identical, but the real space
version turns out to be useful later. Solving equation (\ref{delzeta2}) yields 

\begin{equation}
\bzeta_2=-\bnabla\int\, {d^3a^\prime\over 8\pi} {\bnabla\boldeta_1:
\bnabla\boldeta_1\over |\ba-\bap|} \, . \label{solve}
\end{equation}

\noindent We now write

\begin{equation}
\cL=\cL_2+\cL_4, \label{twofour}
\end{equation}

\noindent (the third order terms vanish) with 
\begin{equation}
\cL_2={\rho\over 2}\int\, d^3a\left(\left|{\partial\boldeta_1\over\partial t}\right|^2 -\valf^2\left|{\partial\boldeta_1\over \partial\al}\right|^2\right), \label{ltran}
\end{equation}

\noindent and

\begin{equation}
\cL_4={\rho\over 2}\int\, d^3a\left(\left|{\partial\bzeta_2\over\partial t}\right|^2 -\valf^2\left|{\partial\bzeta_2\over \partial\al}\right|^2\right). 
\label{llong}
\end{equation}
 
\noindent Variation of the action 

\begin{equation}
{\cal S}\equiv \int\, dt\left\{\cL + \int\, d^3a\, 
P(\bnabla\cdot\boldeta)\right\} \label{action}
\end{equation}

\noindent with respect to $\boldeta$ leads to the (Euler-Lagrange) equation of motion.\footnote{$P$ is a Lagrange multiplier needed to insure
that $\bnabla\cdot\boldeta=0$.}

\subsubsection{3-Wave Interactions In Lagrangian Coordinates}

In this subsection we recover the results of the 3-wave interactions
calculated by NB. Moreover, we demonstrate that they are purely kinematic in
Lagrangian coordinates.
 
To lowest order, the contribution of $\cL_4$ may be ignored.
Using equation~(\ref{ltran}) for $\cL_2$ in the variation
of $S$ in equation~(\ref{action}) results in the following simple, linear equation:
\footnote{We assign first order to $\boldeta$ of unperturbed wave trains.}   

\begin{equation}
\left({\partial^2\over\partial t^2}-\valf^2{\partial^2\over\partial\al^2}
\right)\boldeta_1=0\,,\label{waveeqn}
\end{equation}

\noindent whose general solution is a superposition of wavepackets
traveling in the positive, and  negative $z$ directions:

\begin{equation}
\boldeta_1=\boldeta_1^+(\ba_\perp, \al -\valf t) +
\boldeta_1^-(\ba_\perp, \al +\valf t)\,.\label{superpose}
\end{equation}

NB calculated, in Eulerian coordinates, the lowest order perturbation  
due to a collision between oppositely directed wave packets.
It is a trivial matter to obtain this quantity by using Lagrangian
perturbation theory. To do so, we transform the right side of the 
expression for ${\bf v}$, given in equation~(\ref{vbdef}), into 
Eulerian coordinates. To first order, 

\begin{equation}
{\bf v}_1(\bx, t)={\partial\boldeta_1\over\partial t}(\bx, t)\,,
\label{unpvel}
\end{equation}

\noindent is the velocity field of unperturbed wave packets, and $\boldeta_1$ is 
the corresponding displacement field. To second order, 

\begin{equation}
{\bf v}_2= -(\boldeta_1\cdot\bnabla_x){\bf v}_1+{\partial\bzeta_2
\over\partial t}\,,\label{pertvel}
\end{equation}

\noindent where the subscript `$x$' refers to Eulerian coordinates.\footnote{The
absence of $\partial\boldeta_2/\partial t$ is due to the
vanishing of $\cL_3$.}~ When we substitute equation~(\ref{superpose}) on the right side of equation~(\ref{pertvel}), we obtain three different types of terms; those 
that contain two powers of $\boldeta_1^+$ or $\boldeta_1^-$ have nothing to
do with perturbations induced by collisions. Only the mixed terms 
describe distortions suffered by a wave packet during a collision with an 
oppositely directed wave packet. If 

\begin{equation}
\bDelta_1^{\pm}=\mp[\boldeta_1^\pm(\bx_\perp, +\infty) -
\boldeta_1^\pm(\bx_\perp, -\infty)]\,,\end{equation}

\noindent is the net displacement of field lines due to the 
$+/-$ wave packets, then the asymptotic distortion due to a collision is

\begin{equation}
\Delta{\bf v}_2^\pm= -(\bDelta_1^\mp\cdot\bnabla_x){\bf v}_1^\pm
-\bnabla_x\int\, {d^3x^\prime\over 4\pi} {\bnabla_x(\bDelta_1^\mp):
\bnabla_x{\bf v}_1^\pm\over |\bx-\bx^\prime|}\,.\label{netchange}
\end{equation}

\noindent The above expression is equivalent to equations~(15) and (16) of NB.
The wave packet distortions expressed 
by equation~(\ref{netchange}) are kinematic, as described in \S~2.2.3.
Displacements resulting from any sequence of 
collisions are obtained by summing individual values.\footnote{SG missed these 
distortions because they studied wave packet interactions in which the 
wave packets were localized in $\boldeta$; for these $\bDelta_1^\pm=0$, and 
third order couplings vanish even in Eulerian coordinates. This is 
consistent with the point made in footnote~5 of SG, that resonant coupling coefficients are independent of the variables used. Of course, for this to be 
true, one must be in a regime where perturbation theory is valid, which 
obtains only for wave packets localized in $\boldeta$.}~Note that both terms 
on the right side of equation~(\ref{netchange}) depend on $\bDelta_1^\pm$, 
the net displacement of field lines, which is related to the Fourier 
amplitude of $\boldeta_1^\pm$ with $k_z=0$.

\subsubsection{4-Wave Interactions in Lagrangian Coordinates}

Our principal aim in this subsection is to demonstrate that 4-wave 
interactions obey the distance scaling proposed in equation (\ref{eqdelnL}). 
Our secondary goals are to obtain the spectrum of weak Alfv\'enic turbulence
from a configuration space calculation, and to recover the frequency changing
terms discovered by NB in full MHD.\footnote{NB's discovery was made using
reduced MHD.}

We have shown that the 3--wave interactions of NB arise from kinematic 
perturbations in Eulerian coordinates. Dynamic perturbations require
the interaction of at least 4 waves and are associated with
perturbations in Lagrangian coordinates. We already possess the machinery
necessary to derive these. When $\cL_4$ is included in the action of 
equation~(\ref{action}), the variation with respect to $\boldeta_1$ leads to
Euler--Lagrange equations, which may be thought of as adding third order
terms to the right side of equation~(\ref{waveeqn}):
 
\begin{equation}
\left({\partial^2\over\partial t^2}-\valf^2{\partial^2\over\partial\al^2}
\right)\boldeta_3=(\boldeta_1\cdot\bnabla)
\left({\partial^2\over\partial t^2}-\valf^2{\partial^2\over\partial\al^2}
\right)\bzeta_2 -{\bnabla\tilde{P}_3\over\rho}\,,\label{waveeqnfull}
\end{equation}

\noindent where $\tilde{P}_3$ is determined by requiring that $\bnabla\cdot\boldeta_3=0\,$.

For simplicity, we evaluate the deformation suffered by a wave packet
traveling in the positive $\al$ direction. It proves convenient
to transform to coordinates 

\begin{equation}
\alpha=\al-\valf t, \quad\quad \tau=t. \end{equation}

\noindent As given in equation~(\ref{superpose}), the
unperturbed wave packets in these coordinates have the forms 

\begin{equation}
\boldeta_1^+=\boldeta_1^+(\baperp,\alpha) \quad\quad \boldeta_1^-=\boldeta_1^-(\baperp, \alpha+2\valf \tau). \end{equation}

\noindent The equation of motion for $\boldeta_3$ in the new variables, 
$\tau$ and ${\bf A}\equiv (\baperp,\alpha)$, reads

\begin{equation}
{\partial\over \partial \tau}\left({\partial\over \partial \tau}-2\valf
{\partial\over \partial\alpha}\right)\boldeta_3
=-(\boldeta_1\cdot
\bnabla){\partial\over \partial \tau}\left({\partial\over \partial \tau}-2\valf
{\partial\over \partial\alpha}\right)\bnabla\int\,{d^3A^\prime\over 4\pi}{\bnabla\boldeta_1^-:\bnabla\boldeta_1^+\over |{\bf A} -{\bf A}^\prime|} -{\bnabla\tilde{P}_3\over\rho}\,,\label{dynper}
\end{equation}

\noindent where the primes indicate dummy integration variables, and 
the choice of superscripts applied to $\boldeta_1$ on the right side
of the equation is dictated by the requirement that the
differential operators acting on the integral don't kill it;
one $\boldeta_1^+$ and
one $\boldeta_2^-$ must appear in the integral over $d^3A^\prime$ because
the derivatives $\partial/\partial\tau$ and
$(\partial/\partial\tau-2\valf\partial/\partial\alpha)$ annihilate
$\boldeta_1^+$ and $\boldeta_2^-$, respectively.\footnote{Proof of the second
of these relations requires an integration by parts to transfer
$\partial/\partial\alpha$ acting on $1/|{\bf A} -{\bf A}^\prime|$ to
$\partial/\partial\alpha^\prime$ acting on 
$\bnabla\boldeta_1:\bnabla\boldeta_1$.}~

Further expansion of the right side of equation~(\ref{dynper}), obtained by
writing $\boldeta_1=\boldeta_1^+ +\boldeta_1^-$, results in two terms. 
Rather than carry the cumbersome pressure term through the remainder
of our calculation, we discard it and replace $\boldeta_3$
on the left side of the equation of motion by $\tilde{\boldeta}_3$ to remind us that its longitudinal part must be subtracted off at a later stage: 

\begin{eqnarray}
{\partial\over \partial \tau}\left({\partial\over \partial \tau}-2\valf
{\partial\over \partial\alpha}\right)\tilde{\boldeta}_3
&= &-\left({\partial\over \partial \tau}-2\valf
{\partial\over \partial\alpha}\right)(\boldeta_1^-\cdot
\bnabla){\partial\over \partial \tau}\bnabla\int\,{d^3A^\prime\over 4\pi}{\bnabla\boldeta_1^-:\bnabla\boldeta_1^+\over |{\bf A} -{\bf A}^\prime|}
\nonumber \\
&{}&-{\partial\over \partial \tau}(\boldeta_1^+\cdot
\bnabla)\left({\partial\over \partial \tau}-2\valf
{\partial\over \partial\alpha}\right)\bnabla\int\,{d^3A^\prime\over 4\pi}{\bnabla\boldeta_1^-:\bnabla\boldeta_1^+\over |{\bf A} -{\bf A}^\prime|} 
\,.\label{dynpert}
\end{eqnarray}

\noindent Although we have used coordinates, $(\alpha, \tau)$, moving with the positive
wave packet, we emphasize that the equation of motion (eqn.~[\ref{dynpert}]) 
remains symmetric in $\tilde{\boldeta}_3^+$ and $\tilde{\boldeta}_3^-$: i.e.,
the action of the first term on a $+$($-$) wave packet is identical to that of
the second term on a $-$($+$) wave packet. To determine the perturbations
suffered by 
a $+$ wave packet, we set $\tilde{\boldeta}_3=\tilde{\boldeta}_3^+$ on the 
left side.  

If the first term were the only perturbing interaction, we could peel off
the operator $(\partial/\partial\tau-2\valf\partial/\partial\alpha)$ from
both sides. Integration over time would yield 

\begin{equation}
\Delta\tilde{\boldeta}^+_{3a}=-\int_{-\infty}^{+\infty}d\tau
(\boldeta_1^-\cdot
\bnabla)\bnabla\int\,{d^3A^\prime\over 4\pi}{\bnabla
(\partial\boldeta_1^-/\partial\tau):
\bnabla\boldeta_1^+\over |{\bf A} -{\bf A}^\prime|}
\,,\label{more4}\end{equation}

\noindent which describes the 4--wave interactions (cf. eqn.~[\ref{reso4}])
that form the basis of the weak turbulence theory of SG. 

The third order Eulerian velocity perturbation consists of both kinematic and dynamic terms. An explicit expression follows from equation (\ref{vbdef}):

\begin{eqnarray}
{\bf v}_3(\bx, t) &=& {1\over 2}\boldeta_1\boldeta_1:\bnabla_x\bnabla_x{\partial\boldeta_1\over\partial t}
+\left(\boldeta_1\cdot\bnabla_x\right)\left(\boldeta_1\cdot\bnabla_x\right)
{\partial\boldeta_1\over\partial t} -\left(\boldeta_1\cdot\bnabla_x\right)
{\partial\bzeta_2\over\partial t}
\nonumber \\
&{}& -\left(\bzeta_2\cdot\bnabla_x\right)
{\partial\boldeta_1\over\partial t}+{\partial\bzeta_3\over\partial t}
+{\partial\boldeta_3\over\partial t}\, . \label{v3}
\end{eqnarray}

\noindent The final term is the sole dynamic entry. Each of the other terms is constructed from $\boldeta_1$ without use of the
equation of motion. In particular, $\bzeta_3$ is obtained from equation (\ref{delzeta3})

Let us estimate the fractional distortion suffered by a positive wave packet
after traveling a distance $z\gg \ell$. Among the plethora of kinematic terms, consider only 

\begin{equation}
\Delta{\bf v}_3^+={1\over 2}\Delta\left(\boldeta_1^-\boldeta_1^-\right): \bnabla_x\bnabla_x{\partial\boldeta_1^+\over\partial t}\, . \label{kinone}
\end{equation}
 
\noindent To order of magnitude both this term and the dynamic term 
contribute

\begin{equation}
{\delta^4 v^+_\lambda\over v^+_\lambda}
\sim\left({\ell v_\lambda^-\over\lambda\valf}\right)^2
\left({z\over\ell}\right)\, ,\label{deletap}
\end{equation}

\noindent  the same order of magnitude as the fractional Eulerian distortion
given by equation (\ref{eqdelnL}).\footnote{We draw particular attention
to the linear dependence on $z/\ell$.}

To make contact with the weak 4--wave cascade, we suppose that the velocity spectrum of the negatively directed waves is
cut off at $k_z\sim L^{-1}\ll \ell^{-1}$.\footnote{For simplicity we consider the symmetric situation and drop the $\pm$ signs in the superscripts.} Then
the distortions build up coherently over distance $L$. Over longer
distances, $z\gg L$, they add with random phases implying

\begin{equation}
{\delta^4 v_\lambda\over v_\lambda} \sim \left({\ell v_\lambda\over
\lambda\valf}\right)^2
\left({L\over\ell}\right)\left({z\over L}\right)^{1/2}.
\end{equation}

\noindent Cascade occurs when $\delta^4v_\lambda/v_\lambda\sim 1$; thus

\begin{equation}
N_\lambda\sim\left({\lambda\valf\over\ell v_\lambda}\right)^4
\left({\ell\over L}\right). 
\end{equation}

\noindent But for the extra factor of $(\ell/L)$, this is identical to
equation (\ref{Nfour}), whereupon, provided $N_\lambda\gg 1$, the spectrum of the weak 4--wave cascade, given in equation (\ref{direct}) follows.

Next we return to investigate the second term in equation (\ref{dynpert}).
Imagine that the first term were switched off. Then remove 
a $\partial/\partial\tau$ from both sides leaving the left side in the form 
$(\partial/\partial\tau-2\valf\partial/\partial\alpha)\tilde{\boldeta}^+_{3b}$.
We expect the action of $\partial/\partial\tau$ to be sub--dominant.
Therefore, 

\begin{equation}
\left({\partial\tilde{\boldeta}^+_{3b}\over \partial\alpha}\right)\simeq
-(\boldeta_1^+\cdot
\bnabla)\bnabla\int\,{d^3A^\prime\over 4\pi}{\bnabla\boldeta_1^-:\bnabla
(\partial\boldeta_1^+/\partial\alpha^\prime)
\over |{\bf A} -{\bf A}^\prime|}\,,\end{equation}

\noindent implying that the net change in the $+$ wave packet is

\begin{equation} 
\Delta\left({\partial\tilde{\boldeta}^+_{3b}\over \partial\alpha}\right)\simeq
-(\boldeta_1^+\cdot
\bnabla)\bnabla\int\,{d^2a_\perp^\prime\over 4\pi}
\bigl[\bnabla\boldeta_1^-\bigr ]^\infty_{-\infty}:
\int\, d\alpha^\prime\,{\bnabla(\partial\boldeta_1^+
/\partial\alpha)\over
|{\bf A}-{\bf A}^\prime|}. \label{freqchange}
\end{equation}

We have written equation (\ref{freqchange}) in a form that makes explicit the dependence of the perturbation on only {\sl the net
displacement induced by the $-$ wave packet}.
In other words, the perturbation induced in a wave packet is 
proportional to the amplitudes of the $\kz=0$ Fourier components of  
oppositely directed wave packets: these terms were first discovered by
NB in the limit of reduced--MHD. We may describe the interactions
by the following kind of resonant 4--wave process,

\begin{eqnarray}
\ka &=&\kb + \kc + \kd\,,\nonumber \\
\omega_1 &=& \omega_2 + \omega_3 + \omega_4\,,\label{newreso4}
\end{eqnarray}

\noindent wherein one wave may be induced (by an oppositely directed wave)
to split into two, or two waves could equally well be induced to
combine into one.
Manipulating the resonance relations leads to
$k_{1z}$, $k_{2z}$, $k_{3z}$ positive (say), while $k_{4z}$
approaches zero from below. Thus $\omega_1 = \omega_2 + \omega_3$
allows for changes in the frequencies of wave packets, and
harmonics, as well as sub--harmonics are generated by this process.
These 4--wave interactions 
require the energy spectrum of the negatively directed packets to be 
flat near $k_z=0$; thus they do not arise in weak turbulence.
Are they of importance to intermediate turbulence?  To this end, let us
make an order of magnitude estimate of the perturbation. From
equation~(\ref{freqchange}), the perturbation suffered by the $+$ wave
packet upon traveling a distance $z\sim\valf t$ is

\begin{equation}
\delta\eta_\lambda^+\sim k_\perp^2\,\eta_\lambda^+\eta_\lambda^+
\Delta\eta_\lambda^-\,,\end{equation}

\noindent where $\Delta\eta_\lambda^- \equiv\eta^-(\lambda_\perp, z)
-\eta^-(\lambda_\perp, 0)$.  Using $\eta_\lambda\sim \ell(v_\lambda/v_A)$,

\begin{equation}
\Delta\eta_\lambda^-\sim{v^-_\lambda\over v_A}(\ell z)^{1/2}\,.
\label{Deletam}
\end{equation}

\noindent Thus the fractional distortion of the positively directed wave packet,

\begin{equation}
{\delta v^+_\lambda\over v^+_\lambda}\sim {\delta\eta^+_\lambda\over\eta^+_\lambda}
\sim\left({\ell v_\lambda^+\over\lambda\valf}\right)
\left({\ell v_\lambda^-\over\lambda\valf}\right)
\left({z\over\ell}\right)^{1/2}\,,\label{harmonic}
\end{equation}

\noindent which is smaller than the corresponding quantity in intermediate 
turbulence (set $n=4$ in eqn.~[\ref{eqdelnL}], and compare with the 
above eqn.~[\ref{harmonic}]) by a factor $(z/\ell)^{1/2}$; 
thus harmonic generation is unimportant for intermediate turbulence.\footnote{However, it might play an important role in 
strong turbulence.}

\section{EVOLUTION OF A WEAK PERTURBATION}

Now we address the issue of the physical relevance of the cascade
proposed by SG for weak turbulence. We go on to show that weak
turbulence, intermediate turbulence, and strong turbulence can be
consecutive stages of a single turbulent cascade. 

Imagine exciting shear Alfv\'en waves (isotropically on scale $\ell/L\ll 1$ 
with amplitude $v_\ell/\valf\ll 1$) in a cubical box filled with an 
electrically conducting fluid which is threaded by an unperturbed
magnetic field aligned parallel to the $z$ axis. Let the box have side length
$L$ and assume that its walls are made of an excellent electrical conductor.
Then field line displacements associated with turbulent motions must vanish
at the walls. This boundary condition provides the cutoff of the energy 
spectrum for $k_zL<1$. 

Suppose that

\begin{equation}
\left({v_\ell\over\valf}\right)^2\ll {\ell\over L}. \end{equation}

\noindent Then field lines initially separated by scale $\ell$ maintain this
approximate spacing. This ensures that there are no resonant 3-wave
interactions in the upper part of the cascade and that resonant 4-wave interactions dominate.  From equation (\ref{direct}) we have 

\begin{equation}
{v_\lambda\over\valf}\sim {v_\ell\over\valf}\left({\lambda\over \ell}\right)^{2/3}. \end{equation}

\noindent The weak cascade grades into the intermediate cascade when 

\begin{equation}
\left({v_\lambda\over\valf}\right)^2 \sim {\lambda^2\over \ell L}, 
\end{equation}

\noindent which occurs for 

\begin{equation}
{\lambda_1\over \ell}\sim \left({v_\ell\over\valf}\right)^{3}\left({L\over \ell}\right)^{3/2}. \end{equation}

\noindent For the intermediate cascade

\begin{equation}
{v_\lambda\over\valf}\sim \left({v_\ell\over\valf}\right)^{3/2}\left({L\over \ell}\right)^{1/4}\left({\lambda\over \ell}\right)^{1/2}. \end{equation}

\noindent The intermediate cascade steepens into the strong cascade when

\begin{equation}
{v_\lambda\over\valf}\sim {\lambda\over \ell}, \end{equation}

\noindent which takes place at

\begin{equation}
{\lambda_2\over \ell}\sim  \left({v_\ell\over\valf}\right)^{3}\left({L\over \ell}\right)^{1/2}. \end{equation}

\noindent Within the inertial range of the strong cascade 

\begin{equation}
{v_\lambda\over\valf}\sim \left({v_\ell\over v_A}\right)^2\left({L\over \ell}\right)^{1/3}\left({\lambda\over \ell}\right)^{1/3}. \end{equation}

The complete three dimensional, inertial range spectrum is given by

\begin{eqnarray}
E(k_z, \kperp)\sim \ell^3\vl^2\left\{\begin{array}{lll}(\kperp\ell)^{-10/3},  & \mbox{if ($L^{-1}<k_z<\ell^{-1}$,
and $\ell^{-1}<\kperp<\lambda_1^{-1}$);}\\
(\kperp\ell)^{-3},&\mbox{if ($L^{-1}<k_z<\ell^{-1}$,
and $\lambda_1^{-1}<\kperp<\lambda_2^{-1}$);}\\
(\kperp\ell)^{-8/3},&\mbox{if ($L^{-1}<k_z<\ell^{-1}$,
and $\kperp >\lambda_2^{-1}\,$).}\end{array}\right.\label{evolspec}
\end{eqnarray}

\noindent There are a couple of points worth noting in connection with the above
combined cascade. The intermediate cascade is confined to $\lambda_2\lesssim \lambda\lesssim \lambda_1$ where the ratio  

\begin{equation}
{\cal R}\equiv {\lambda_2\over \lambda_1}\sim {\ell\over L},
\end{equation}

\noindent is independent of $v_\ell/\valf$. As ${\cal R}\to 1$ from below, the
inertial range of this cascade shrinks to zero exposing a direct
transition between the weak and strong cascades. This is the transition
discussed in SG and GS.

\section{DISCUSSION}

In a recent paper, Ng \& Bhattacharjee~(1996) claim that (i) in weak
MHD turbulence, 3--wave interactions between oppositely directed
wave packets are non zero if the $k_z=0$ components are non zero; (ii) 3--wave interactions dominate over 4--wave interactions. We hope to have persuaded the 
reader that (i) is correct so long as ``weak'' is altered to ``intermediate'', and (ii) is true only for individual collisions 
between small amplitude wave packets. However, NB deserve full
credit for demonstrating the importance of 3--wave interactions, as 
well as discovering the frequency--changing terms in 4--wave interactions.
Both are a consequence of a non zero net displacement of field lines,
due to the perturbations of wave packets that have localized velocity and magnetic field perturbations. 
 
Intermediate turbulence shares
with weak turbulence the property that wave packets are long lived; interaction
times are much longer than the wave periods. However, the distinguishing feature
is that perturbation theory in not applicable to intermediate turbulence; this 
should be clear from our demonstration that interactions of all orders have the same strength. Of course, during individual collisions, 3--wave interactions  
dominate over all higher order interactions. However, as described in \S~2.4, 
for wave packets
localized in velocity and magnetic fields, but not in the net displacement of
field lines, subsequent collisions are correlated; this makes all n--wave 
interactions contribute equally to intermediate turbulence. If Iroshnikov and 
Kraichnan had performed their heuristic estimates, taking account of the fact 
that there is no parallel cascade for long lived wave packets, they would have 
found the spectrum of the intermediate cascade. This is one case in which
the assumption of isotropy (usually innocuous) is misleading;
the anisotropic 
intermediate cascade strengthens on small spatial scales, whereas the 
isotropic IK cascade 
weakens.  

We have devoted attention almost exclusively to {\sl shear} Alfv\'en 
($\sa$ for brevity) waves,
ignoring the dynamics of {\sl pseudo} Alfv\'en ($\pa$ for brevity)
waves.\footnote{The only
exception is the proof that there are no 3--wave couplings for either type
of Alfv\'en wave, in the Lagrangian perturbation theory of SG.}~ A generic 
excitation may be expected to put power equally into both kinds of waves. 
Shouldn't we then study the interactions of the $\pa$ waves among themselves,
as well as with $\sa$ waves? Will this modify the cascades we derived for 
the $\sa$ waves? It turns out that the $\pa$ waves are {\sl slaved} to the 
$\sa$ waves. The reason is as follows. Suppose that we were following the 
distortions suffered by a positively directed wave packet due to other, 
negatively directed ones. The nonlinear interactions  are given, to order
of magnitude, by some power of $({\bf v}^{-}\cdot\bnabla){\bf v}^{+}\sim
({\bf k}\cdot{\bf v}^{-}){\bf v}^{+}$. Because the cascades in MHD turbulence
are anisotropic, we expect that $\kperp\gg k_z$ in the inertial range. We note that 
the polarization of an $\sa$ wave is essentially along $\hat{z}{\bf\times}
{\bf k}_\perp$, and  that of a $\pa$ wave is along $\hat{z}$, and hence estimate
that the ``operator'' ${\bf k}\cdot{\bf v}\sim (\kperp v_\sa) + (k_z v_\pa)\sim
(\kperp v_\sa)$.\footnote{If the amplitude of the 
$\pa$ wave is very much larger than the $\sa$ wave, we could imagine that
$(\kperp v_\sa)<(k_z v_\pa)$, but this is an unlikely situation. However, 
a detailed analysis rules out this possibility in cases where both waves
have comparable amplitude on the outer scale.}~ Thus the $\pa$ waves are slaved
to the $\sa$ waves; in a later paper, we will derive kinetic equations, and 
demonstrate that the spectrum of the $\pa$ waves is identical to that of the $\sa$
waves for weak, intermediate, and strong turbulence.

How might an energy spectrum with a cutoff below $k_zL\sim 1$ arise? 
Two related possibilities come to mind. The first involves a thought experiment
that could be realized computationally; we described this in \S~4. 
A more natural setting might be the atmosphere of a
massive star which possesses a strong external dipole field.\footnote{Stars 
with masses in excess of a few 
solar masses have radiative atmospheres with little mass motion.}~  The energy spectrum of waves in the stellar magnetosphere would be cutoff on scales longer than the length of the
flux tubes that link the northern and southern magnetic hemispheres. 
The cut off at small $\kz$ necessary for weak turbulence 
makes it, in general, less applicable than intermediate turbulence.
However, the limited inertial ranges of both weak and intermediate 
cascades suggests that neither is likely to find much application
in nature.  
The critically balanced cascade, proposed by Goldreich \& Sridhar~(1995) 
for strong MHD turbulence, remains as the  most likely candidate for 
interstellar turbulence. 

\acknowledgments

Financial support for this research was provided by NSF grant 94-14232.
PG benefited from visits to the Institute for Advanced Study and the Canadian Institute for Theoretical Astrophysics. We thank Omer Blaes for comments which improved the clarity of the manuscript.

\end{document}